\documentstyle[prl,aps,mypsfig2,twocolumn]{revtex}

\begin{document}

\preprint{}

\title{Realization of logically labeled effective pure states for bulk quantum computation}

\author{Lieven M.K. Vandersypen{$^{1,2,\dagger}$}, Costantino S. Yannoni{$^2$}, Mark H. Sherwood{$^2$}, Isaac L. Chuang{$^2$}}
  
\address{\vspace*{1.2ex}
	{$^1$ Solid State  and Photonics Laboratory, Stanford University, 
		  Stanford, CA 94305-4075}\\[1.2ex]
 	{$^2$ IBM Almaden Research Center,  San Jose, CA 95120}\\[1.2ex]
}

\date{\today \vspace*{-.3cm}}

\maketitle

\def\>{\rangle}
\def\be{\begin{equation}}
\def\ee{\end{equation}}
\def\bea{\begin{eqnarray}}
\def\eea{\end{eqnarray}}
\newcommand{\ket}[1]{\mbox{$|#1\rangle$}}
\newcommand{\bra}[1]{\mbox{$\langle #1|$}}
\newcommand{\mypsfg}[2]{\psfig{file=#1,#2}}


\begin{abstract}

\vspace*{-5ex}

We report the first use of "logical labeling" to perform a quantum computation with a room-temperature bulk system. This method entails the selection of a subsystem which behaves as if it were at zero temperature --- except for a decrease in signal strength --- conditioned upon the state of the remaining system. No averaging over differently prepared molecules is required. In order to test this concept, we execute a quantum search algorithm in a subspace of two nuclear spins, labeled by a third spin, using solution nuclear magnetic resonance (NMR), and employing a novel choice of reference frame to uncouple nuclei.\\

\end{abstract}


The extraordinary potential of quantum
algorithms~\cite{Deutsch85,Shor94,Grover97,Divincenzo95,Ekert96} to solve
certain problems in fewer steps than the best known classical algorithms,
has recently been highlighted by the experimental demonstration of
elementary quantum
computations~\cite{Chuang98a,Chuang98b,Jones98a,Jones98b,Cory98,Linden98}.
These experiments made use of nuclear spins, subject to a magnetic field, as
quantum bits (qubits: 2-level quantum systems). Even though traditional
quantum algorithms assume that all qubits are necessarily initialized in a
pure state, typically the ground state, the experiments were carried out at
room temperature and the spins were in a highly mixed state.

The use of room temperature nuclear spins for quantum computing relies on
the ideas of Gershenfeld and Chuang~\cite{Gershenfeld97} and Cory, Fahmy and
Havel~\cite{Cory97} to convert the mixed thermally equilibrated state into
an ``effective pure'' state, at the expense of a signal decay, exponential the number of qubits. The density matrix of such a state is diagonal
and proportional to the identity matrix, except that one of the diagonal
elements is different from the others. Because states proportional to the
identity matrix are not observable in NMR spectroscopy and do not transform
under unitary operations, an effective pure state gives the signal and has
the dynamical behavior of a pure state. Until now, the signal of an
effective pure state has been extracted by adding the signals from
differently prepared molecules, either in different locations using magnetic
field gradients~\cite{Cory97} or by time-sequential
experiments~\cite{Knill97}.

Logical labeling offers a fundamentally different approach for creating effective pure states: a submanifold spanned by $2^k$ states is selected from the Hilbert space of $n$ coupled spins, by conditioning upon the state of $n-k$ ancilla spins. Gershenfeld and Chuang~\cite{Gershenfeld97} predicted that the structure in the density matrix at equilibrium can be exploited to make this subsystem behave as if it were in a pure state, even if the entire system is at room temperature. This concept of embedding was previously used to observe Berry's phase in NMR spectroscopy~\cite{Suter86}.


In this Letter, we report the experimental implementation of logical labeling by embedding an effective pure 2-qubit subspace within a 3-spin molecule. Control over the dynamical behavior of the two logically labeled qubits is facilitated by the ``uncoupling frame'', a novel method to remove the effect of undesired couplings on the spin dynamics within a subspace. We execute Grover's quantum search algorithm as a test of logical labeling and the uncoupling frame, and study the preservation of the effective pure state after many operations.


The Hamiltonian of weakly coupled nuclear spins in a molecule in solution and subject to a strong magnetic field, is well approximated by~\cite{Abragam} ($\hbar = 1$)
\be
{\cal H} = - \sum_{i} \omega_i I_{z i}
 	   + \sum_{i<j} 2\pi J_{i j} I_{z i} I_{z j}
	   + {\cal H}_{env} ,
\ee
where $2 I_{z i}$ is the $\hat{z}$ Pauli operator and $\omega_i$ is the Larmour frequency of spin $i$. $J_{i j}$ is the strength of the scalar spin-spin coupling and ${\cal H}_{env}$ represents coupling to the environment, which causes decoherence. For a 3-spin system at typical magnetic field strengths, the Boltzmann distribution describing the thermal equilibrium populations of the states $\{ \ket{000},$ $\ket{001},$ $\ket{010},$ $\ket{011},$ $\ket{100},$ $\ket{101},$ $\ket{110},$ $\ket{111} \}$ (where 0 and 1 represent a spin in the ground and excited state respectively) is dominated by the Zeeman energy shift $\pm \hbar \omega_i /2$ and, for homonuclear systems ($\omega_i \approx \omega, i=1,2,3$), is well approximated by 
\be
\frac{1}{2^3} \{\,1\hspace*{-.5ex}+\hspace*{-.2ex}3a, 1\hspace*{-.5ex}+\hspace*{-.2ex}a, 1\hspace*{-.5ex}+\hspace*{-.2ex}a, 1\hspace*{-.5ex}-\hspace*{-.2ex}a, 1\hspace*{-.5ex}+\hspace*{-.2ex}a, 1\hspace*{-.5ex}-\hspace*{-.2ex}a, 1\hspace*{-.5ex}-\hspace*{-.2ex}a, 1\hspace*{-.5ex}-\hspace*{-.2ex}3a \,\},
\label{eq:denmat_eq}
\ee
where $a$ = $\hbar \omega / 2k_B T <<$ 1, with $k_B T$ the thermal energy. The thermal equilibrium state is thus highly mixed. However, the observable signal from the subspace spanned by the states $\ket{000},$ $\ket{011},$ $\ket{101}$ and $\ket{110}$ is the same as the signal from a system in the state $\ket{000}$, to within a scaling factor. This subspace is thus in the effective pure state $\ket{000}$.

In order to simplify subsequent logical operations and to separate the signals of the effective pure subspace and its complement, the populations can be rearranged by a sequence of 1 and 2-qubit unitary operations to obtain
\be
\frac{1}{2^3} \{\,1\hspace*{-.5ex}+\hspace*{-.2ex}3a, 1\hspace*{-.5ex}-\hspace*{-.2ex}a, 1\hspace*{-.5ex}-\hspace*{-.2ex}a, 1\hspace*{-.5ex}-\hspace*{-.2ex}a, 1\hspace*{-.5ex}+\hspace*{-.2ex}a, 1\hspace*{-.5ex}+\hspace*{-.2ex}a, 1\hspace*{-.5ex}+\hspace*{-.2ex}a, 1\hspace*{-.5ex}-\hspace*{-.2ex}3a \,\}.
\label{eq:denmat_lab}
\ee
Now the subspace $\{ \ket{000},$ $\ket{001},$ $\ket{010},$ $\ket{011} \}$ is in an effective pure state. This subspace corresponds to the spins $B$ and $C$ conditioned on or {\em labeled} by the state of the first spin, $A$, being $\ket{0}$ (we will call this the $\ket{0}_A$ subspace). This procedure is called logical labeling~\cite{Gershenfeld97} and, combined with removal of coupling to spin $A$ for the remainder of the pulse sequence, allows 2-qubit quantum computations involving only the spins $B$ and $C$.

\begin{figure}[t]
\vspace*{-2ex}
\begin{center}
\mbox{\psfig{file=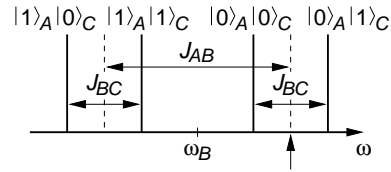,width=2in}}
\end{center}
\vspace*{-1.5ex}
\caption{Spectrum of the transitions of $B$, with the states of the other two spins as indicated. With respect to a reference frame rotating at $\omega_B / 2\pi$ $- J_{AB}/2$ (indicated by an arrow), spins $B$ which see a spin $A$ in $\ket{0}$, evolve under $J_{BC}$ only and are thus {\em uncoupled} from $A$. $J_{AB}$ does affect $B$'s evolution in the $\ket{1}_A$ subspace, but the signal of this subspace does not interfere with that of the $\ket{0}_A$ subspace. Similarly, $C$'s rotating frame must be moved to $\omega_C/2\pi - J_{AC}/2$. A separate channel was used for $B$ and $C$. We note that the uncoupling frame is somewhat related to selective decoupling~\protect\cite{Maher61}, a technique to determine the relative sign of $J$-couplings by moving the reference frame of one nucleus (not several, as here) to the center of a submanifold.}
\label{fig:uncoupling_frame}
\end{figure}

We selected bromotrifluoroethylene dissolved in deuterated acetone (10
mol$\%$) as the central molecule in our experiments, because the spin-1/2
$^{19}$F nuclei have large $J$-couplings and chemical shifts, as well as long
coherence times, which makes it suitable for quantum
computation~\cite{Divincenzo95,Gershenfeld97}. The $^{12}$C nuclei are
non-magnetic and the interaction of the spin-3/2 Br nucleus with the fluorine
spins is averaged out due to fast Br relaxation~\cite{Abragam}. The $^{19}$F
Larmour frequencies are $\approx$ 470 MHz (at 11.7 T) and the spectrum is
first order, consisting of three well-separated quadruplets (as in
Fig.~\ref{fig:uncoupling_frame}), with $\omega_A - \omega_B / 2\pi \approx$
13.2 kHz and $\omega_C - \omega_A / 2\pi \approx$ 9.5 kHz. The coupling
constants are measured to be $J_{AB} = -122.1$ Hz, $J_{AC} = 75.0$ Hz and
$J_{BC}= 53.8$ Hz (see also~\cite{Elleman62}). Single-spin rotations are
realized by applying a radio-frequency (RF) magnetic field, oscillating in the
$\hat{x}-\hat{y}$ plane. To simultaneously address the four lines in one
quadruplet without affecting the other two quadruplets, the envelope of the RF
pulses is Gaussian-shaped (more complex shapes could improve the results) and
the RF power is adjusted to obtain pulses of $\approx 300 \mu$s. Experiments
were performed at IBM using an Oxford Instruments
wide-bore magnet and a Varian $^{\sf UNITY}${\sl Inova} spectrometer with a
Nalorac triple resonance (H-F-X) probe.

\begin{figure}
\begin{center}
\mbox{\psfig{file=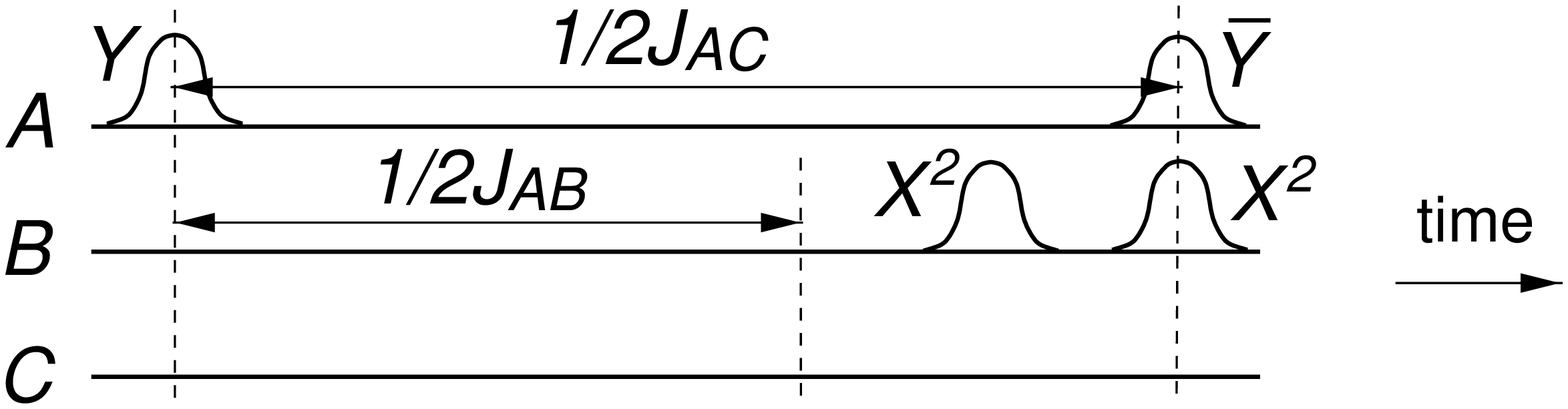,width=2.5in}}
\end{center}
\vspace*{-1ex}
\caption{The logical labeling pulse sequence.
$Y$ and $\bar{Y}$ denote a $90 ^{\circ}$ rotation about the $\hat{y}$ and $-\hat{y}$ axis (right hand rule). $X^2$ are 180$^{\circ}$ rotations. 
}
\label{fig:sequence}
\end{figure}

The logical labeling step is implemented as follows. Comparison of Eq.~\ref{eq:denmat_eq} with Eq.~\ref{eq:denmat_lab} shows that the populations of the states $\ket{001} \leftrightarrow \ket{101}$ and $\ket{010} \leftrightarrow \ket{110}$ must be interchanged, while the remaining four populations must be unaffected. This requires a CNOT$_{BA}$ and a CNOT$_{CA}$, where CNOT$_{ij}$ represents a controlled-NOT which flips $j$ if and only if $i$ is in $\ket{1}$. Furthermore, the two CNOT gates commute and can thus be executed simultaneously, as shown in Fig.~\ref{fig:sequence}.

\vspace*{-0.5ex}

\begin{figure}[htbp]
\begin{center}
\mbox{\psfig{file=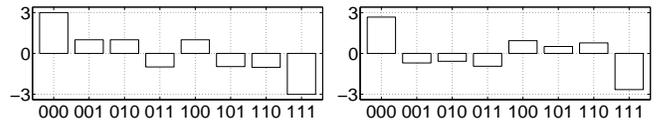,width=3.4in}}
\end{center}
\vspace*{-.3cm}
\caption{Experimentally determined populations (in arbitrary units, and relative to the average) of the states $\ket{000}, \ldots ,\ket{111}$ (Left) in thermal equilibrium and (Right) after logical labeling. The populations were determined by partial state tomography~\protect\cite{Chuang97e}.}
\label{fig:labeling_result}
\end{figure}

This sequence implements the CNOT gates only up to single-spin $Z$ rotations, which suffices for a diagonal initial state as used here. Furthermore, the $I_{zB}$, $I_{zC}$ and $2 I_{zB} I_{zC}$ terms in the Hamiltonian have no effect, since $B$ and $C$ remain along $\pm \hat{z}$. $I_{zA}$ can be ignored because the pulses were applied in a reference frame in resonance with each spin. Fig.~\ref{fig:labeling_result} shows the measured populations of the eight basis states, before and after the sequence of Fig.~\ref{fig:sequence}. The results agree with the theoretical predictions of Eqs.~\ref{eq:denmat_eq} and~\ref{eq:denmat_lab}.

\vspace*{-0.5ex}


The effective pure state prepared with logical labeling must remain pure throughout any subsequent computation. This requires that while a computation is carried out using $B$ and $C$, $A$ must ``do nothing'', which is non-trivial in a system of coupled spins~\cite{Linden98b}: the effect of $J_{AB}$ and $J_{AC}$ must be removed. This could be done by using two refocusing $X^2_A$ pulses during every logical operation between $B$ and $C$. We have devised a different method, which exploits the fact that it suffices to remove the effect of $J_{AB}$ and $J_{AC}$ {\em within the $\ket{0}_A$ subspace}. This uncoupling frame method requires no pulses at all and is described in Fig.~\ref{fig:uncoupling_frame}.


 Mathematically, the transformation of spin $i$'s state from the laboratory frame to a reference frame rotating at $\omega_i + \Delta_i$ (with $\Delta_i$ the resonance off-set frequency), is described by $\ket{\psi} =$ $U \ket{\psi'} =$ ${\rm exp}[i (\omega_i + \Delta_i) I_{z i} t] \ket{\psi'}$, where $\ket{\psi'}$ is the state in the new reference frame. After substituting $\ket{\psi}$ by $U \ket{\psi'}$ in the Schroedinger equation, and using the commutation rules between $I_x$, $I_y$ and $I_z$, it follows that ${\cal H}' = {\cal H} + \hbar (\omega_i + \Delta_i) I_{z i}$~\cite{Abragam}. Then, choosing $\Delta_B = -\pi J_{AB}$ and $\Delta_C = - \pi J_{AC}$ gives

\begin{figure}[t]
\vspace*{-0.8cm}
\begin{center}
\mbox{\psfig{file=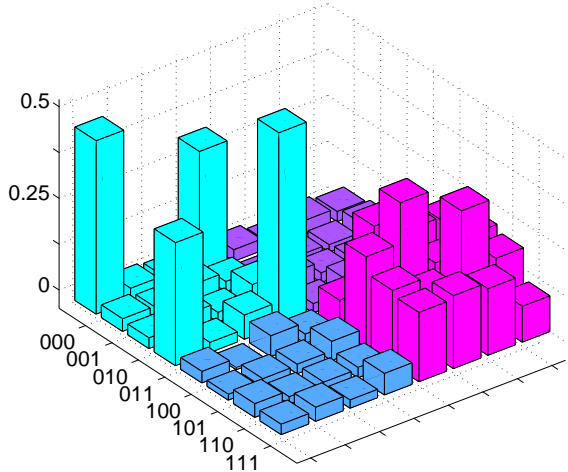,width=3in}}
\end{center}
\vspace*{-1.5ex}
\caption{Normalized experimental deviation density matrix (with the diagonal shifted to obtain unit trace for the effective pure state), shown in absolute value. The entries in the second quadrant are very small, which means that the $\ket{0}_A$ and $\ket{1}_A$ subspaces are uncoupled. The deviation density matrix is measured using quantum state tomography~\protect\cite{Chuang97e}; a series of 27 consecutive experiments with different sets of read-out pulses (see below) permits to determine all the components of the deviation from identity of the density matrix. The relative error in the state is $\parallel\rho_{\rm exp} - \rho_{\rm th}\parallel$/$\parallel\rho_{\rm th}\parallel$ = 19 $\%$.}
\label{fig:labeled_epr}
\end{figure}

\vspace*{-3ex}
$$
\hspace*{-16ex}
{\cal H}' =  2\pi \, [ \; (\mbox{\large \bf 1}/2+I_{zA}) \; J_{BC} I_{zB} I_{zC}
\vspace*{-2ex}
$$
$$
	  + \; (\mbox{\large \bf 1}/2- I_{zA}) \; (J_{BC} I_{zB} I_{zC} - J_{AB} I_{zB} - J_{AC} I_{zC}) \; ] \; ,
\label{eq:frame_hamiltonian}
$$
where {\large \bf 1} is the identity matrix. The first (second) term in the
expression of $\cal{H}'$ acts exclusively on the $\ket{0}_A$ ($\ket{1}_A$)
subspace. Any state within the $\ket{0}_A$ subspace evolves only under $J_{BC}
I_{zB} I_{zC}$ and will remain within the $\ket{0}_A$ subspace. Within the
$\ket{0}_A$ subspace and using the uncoupling reference frame, $B$ and $C$ are
thus uncoupled from $A$. We experimentally confirm this by reconstructing the
deviation density matrix~\cite{Gershenfeld97} of the 3-spin system after
creating the state $(\ket{00} + \ket{11})/\sqrt{2}$ in the $\ket{0}_A$
subspace. Fig.~\ref{fig:labeled_epr} shows the $\ket{0}_A$ 
subsystem in the predicted effective pure state and uncoupled from the 
$\ket{1}_A$ subspace.


The motivation for creating effective pure states is to provide a suitable
initial state for quantum computation, and as a test of this, we implement
Grover's quantum search algorithm~\cite{Grover97} on the two logically labeled
spins, using the uncoupling frame. With this algorithm, one can find the
unique but unknown $x_0$ which satisfies $f(x_0)=1$, in ${\cal O}(\sqrt{N})$
queries of $f(x)$, where $N$ is the number of possible entries $x$. Searching
on a classical machine would require ${\cal O}(N)$ queries of $f(x)$. For
$N=4$, implemented here,  a quantum computer finds $x_0$ in just one
query, compared to on average 2.25 queries classically.


We use a pulse sequence similar to the one described in~\cite{Chuang98a}. First, $Y_B Y_C$ rotates both spins from $\ket{00}$ into $(\ket{00}+\ket{01}+\ket{10}+\ket{11})/2$, an equal superposition of the four possible inputs. The amplitude of the $\ket{x_0}$ term is then amplified in two steps~\cite{Grover97}. First, one of four functions $f_{x_0}(x)$ is evaluated, flipping the sign of the $\ket{x_0}$ term. This is done by one of four conditional phase flips $Y_B Y_C - \Phi_B \Theta_C - \bar{Y}_B \bar{Y}_C - 1/2J_{BC}$, where $\Phi = X$ for $f_{00}$ and $f_{10}$ and $\Phi = \bar{X}$ for $f_{01}$ and $f_{11}$. $\Theta = X$ for $f_{00}$ and $f_{01}$ and $\Theta = \bar{X}$ for $f_{10}$ and $f_{11}$. Second, inversion about the average~\cite{Grover97} is implemented by a Hadamard gate (a $\pi$ rotation about an axis halfway between $\hat{x}$ and $\hat{z}$) on both spins, followed by the conditional phase flip corresponding to $f_{00}$, and another Hadamard gate. The pulse sequence for this inversion step can be reduced to $X_B X_C - Y_B Y_C - 1 / 2 J_{BC} - \bar{Y}_B \bar{Y}_C$. The entire sequence for Grover's algorithm takes approximately 20 ms and the labeling step takes about 7 ms. The coherence time for the three $^{19}$F spins, expressed as the measured transverse relaxation time constant T$_2$ $\approx$ 4-8 s, is sufficiently long for coherence to be maintained throughout the labeling and computation operations.

\begin{figure}[htbp]
\begin{center}
\mbox{\psfig{file=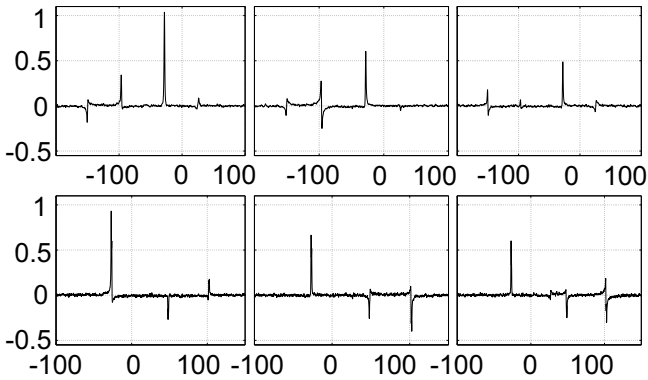,width=3.4in}}
\end{center}
\vspace*{-1.5ex}
\caption{Real part of experimental spectra (frequencies in the uncoupling reference frame) for spin $B$ (Top) and $C$ (Bottom), after executing Grover's algorithm 1 (Left), 19 (Center) and 37 (Right) times, with $\ket{x_0} = \ket{00}$. The $\ket{0}_A$ subspace corresponds to the spectral lines at $\pm J_{BC}/2 = \pm 26.9$ Hz. Ideally, the line at -26.9 Hz is positive and absorptive, with unit amplitude, while the line at +26.9 Hz is zero. Even after 37 iterations, $x_0$ can be unambiguously determined.}
\label{fig:iterations}
\end{figure}

The theoretical prediction is that the state of $BC$ after completion
of the algorithm is the effective pure state $\ket{x_0}$, which can be
determined by a measurement of the spin states of $B$ and $C$. This is
done using a read-out pulse for each spin, which rotates the spin back
into the $\hat{x}-\hat{y}$ plane so that it generates a rotating
magnetic field and induces an oscillating voltage in a pick-up
coil. The voltage is recorded with a phase sensitive detector, and
Fourier transformed. The phase of the receiver is set to obtain a
positive (negative) absorption line for a spin that was $\ket{0}$
($\ket{1}$) before the read-out pulse, and is used to determine
$\ket{x_0}$ (see
Fig.~\ref{fig:uncoupling_frame}). The experimental spectra
(Fig.~\ref{fig:iterations}, Left) as well as the deviation density
matrices of the logically labeled subspace before, during and after
the computation, confirm that the state remains an effective pure
state throughout the computation and that the final state is
$\ket{x_0}$.


An interesting question is how many logical operations can be executed while preserving the effective pure character of the spins, and further, how quickly errors accumulate during longer pulse sequences. We study this by iterating the conditional flip and inversion steps in Grover's algorithm, which ideally gives rise to a periodic pattern: for $N=4$, the amplitude of the $x_0$ term is expected to be 1 after 1 iteration, and again after $4, 7, \ldots$ iterations~\cite{Grover97}. Fig.~\ref{fig:iterations} demonstrates the expected periodic behavior in the output state in experiments with up to 37 iterations, which requires 448 pulses and takes about 700 ms.  In addition to decoherence, errors mainly arise from imperfections in the pulses and coupled evolution during the 300 $\mu$s pulses (reduction of $\tau_1$ and $\tau_2$ partially compensated for this effect). Spectra of the quality of Fig.~\ref{fig:iterations} were obtained by choosing a particular implementation of the composite $\hat{z}$ rotation in the phase flip step, such that the errors it introduces partially cancel with the errors of the inversion step. For example, while $Y X \bar{Y}$, $\bar{Y} \bar{X}$ Y, $X \bar{Y} \bar{X}$ and $\bar{X} Y X$ are all mathematically equivalent, the errors may in practice add up or cancel out with the errors from previous or subsequent pulses. Clearly, a general optimization procedure will be very helpful for designing effective pulse sequences in future experiments involving more qubits.


We attribute the preservation of effective pure states during long pulse sequences in part to the use of the  uncoupling frame, which provides an elegant and simple alternative to refocusing schemes involving $\pi$-pulses. Multiple couplings with ancillae spins can be neutralized simply by moving the carrier frequencies of the computation spins by the appropriate $\sum \pm J/2$. In contrast, refocusing pulses, would have to be applied during every single evolution interval and their complexity rapidly increases as more $J$-couplings are to be refocused~\cite{Linden98b}. This technique may find application in future experiments using logical labeling, as well as in quantum error detection experiments~\cite{Leung98}, where the computation must only proceed within the subspace labeled error-free by the ancillae. However, refocusing pulses are still required when a coupling must be removed over an entire system rather than in a subspace only.


The present experiments demonstrate that a $k$-qubit room temperature system can behave as if it were very cold, except that the signal strength decreases exponentially with the number of qubits, when it is properly embedded in an $n$-spin system. The subspace dimension is limited by the number of equally populated states, which is $C_n^{n/2}$ $= n!/(n/2)!^2$ in a homonuclear system, giving $k = {\rm log}_2(1+C_n^{n/2})$. Thus for large $n$, $k/n$ tends to 1 ($n$=40 for $k$=37). For heteronuclear spin systems, the analysis is more complex and $k/n$ is generally smaller, but the number $k$ of cold qubits that logical labeling can extract from $n$ hot spins, as well as the number of operations required for logical labeling, still scale favorably.


We would like to thank N. Gershenfeld, A. Pines, D. Leung and X. Zhou for useful discussion, and J. S. Harris and N. Amer for their support. L.V. gratefully acknowledges a Yansouni Family Stanford Graduate Fellowship. This work was supported by the DARPA Ultra-scale Program under contract DAAG55-97-1-0341.\\

{$^{\dagger}$ \footnotesize email: lieven@snow.stanford.edu}


\end{document}